\documentclass[preprint,aps,prb,superscriptaddress]{revtex4}
\usepackage{multirow}
\usepackage{graphicx}% Include figure files
\usepackage{dcolumn}% Align table columns on decimal point
\usepackage{bm}% bold math

\usepackage{titlesec}
\titleformat*{\section}{\flushleft \bf \large}
\titleformat*{\subsection}{\flushleft \bf}
\titleformat*{\subsubsection}{\flushleft}
\begin{document}

\title{
$s$-$d$ electronic interactions induced H$_2$ dissociation on the
$\gamma$-U(100) surface and influences of niobium doping}

\author{Yu Yang}
\affiliation{LCP, Institute of Applied Physics and Computational
Mathematics, P.O. Box 8009, Beijing 100088, People's Republic of
China}%
\author{Ping Zhang}
\thanks{To whom correspondence should be
addressed. E-mail: zhang\_ping@iapcm.ac.cn (P.Z.)}%
\affiliation{LCP, Institute of Applied Physics and Computational
Mathematics, P.O. Box 8009, Beijing 100088, People's Republic of
China}%
\author{Peng Shi}
\affiliation{National Key Laboratory for Surface Physics and Chemistry, Mianyang 621907, China}%
\author{Xiaolin Wang}
\affiliation{China Academy of Engineering Physics, Mianyang 621900, China}%

\begin{abstract}

The dissociation of hydrogen molecules on the $\gamma$-U(100)
surface is systematically studied with the density functional theory
method. Through potential energy surface calculations, we find that
hydrogen molecules can dissociate without any barriers on the clean
$\gamma$-U(100) surface. After careful electronic analysis, it is
found that charge transfer between the hydrogen $s$ and uranium $d$
electronic states causes the dissociation, which is quite different
from the dissociation of hydrogen molecules on other actinide metal
surfaces. Considering that doping of 3$d$ transition metal atoms can
stabilize the $\gamma$ phase of U, we also study the influences of
Nb-doping on the hydrogen dissociation process. We find that the
3$d$ electronic states of Nb also take part in the hybridization
with hydrogen $s$ electronic states, which leads to the result that
hydrogen molecules also dissociate without any energy barriers on
the doped U surface. In addition, the free electronic energy lowers
down more quickly for a hydrogen molecule approaching the doped U
surface.

\end{abstract}

\maketitle

\section{Introduction}

During the past decades, many theoretical and experimental studies
have been carried out on the initial processes of surface
hydrogenation or oxidation reactions
\cite{King88,Darling95,Brune92,Brune93,Ma07,Honkala00,Behler05,Behler07,Grob98,Wetzig01,Kroes02,Mitsui03,Nieto06,Grob07,Yang08,Yang09,Yang10}.
The theoretical researches include calculating the potential energy
surfaces for H$_2$ or O$_2$ molecules on solid surfaces, and
revealing the adsorption and dissociation mechanisms of them
\cite{Honkala00,Behler05,Behler07,Grob98,Wetzig01,Kroes02,Mitsui03,Nieto06,Grob07,Yang08,Yang09,Yang10}.
In the experimental aspect, ultrahigh vacuum experiments are
designed to detect the atomic structures after very few numbers of
molecules reacting with solid surfaces \cite{Brune92,Brune93,Ma07}.
Actinides, as a group of radioactive, toxic and rare materials
\cite{Dholabhai07}, have their difficulties to be prepared for such
experimental studies. On the other hand, they play important roles
in advanced nuclear fuel cycles, and surface hydrogenation/oxidation
are the main corrosion mechanisms that fail their storage
\cite{Nie08,Condon73,Condon75,Kirkpatrick81,DeMint00}. Hence,
theoretical studies are crucial for understanding the detailed
surface corrosion mechanisms in the presence of environmental gases
for these high-$Z$ elements. Moreover, these studies may also help
to address the environmental consequences of nuclear materials.

Among the actinide elements, uranium (U) is the heaviest naturally
occurring one, which occupies a central position in the early
actinide series \cite{Dholabhai07,Huda05}. Due to its important role
in nuclear reactors, U is quite familiar to people
\cite{Dholabhai07}. In the atmosphere, U and its alloys are ready to
be oxidized to form U oxide layers, which subsequently break down
through a hydrogenation process. The hydriding reaction proceeds by
surface nucleation and growth of hydride nuclei which spread over
the sample surfaces \cite{Brill98,Shamir09,Shamir10}. It has been
revealed that the hydride nuclei form beneath the oxide layers, and
most of them are capable of penetrating through the oxide layers
above them to grow over time until they consume the sample surface
\cite{Shamir10,Bloch01}. Based on these knowledges, one can see that
the electronic interaction with hydrogen molecules is critically
important for the corrosion of U surfaces. Pure uranium crystallizes
into several structures, the orthorhombic $\alpha$ phase with four
molecules per unit cell at ambient conditions, followed by the
body-centered tetragonal $\beta$ (bct) phase at 940 K and then the
body-centered cubic $\gamma$ (bcc) phase at 1050 K at ambient
pressure \cite{Dholabhai07,Huda05,Koelling73,Young91,Kurihara04}.
Moreover, the high temperature $\gamma$ phase can be studied at
normal temperatures by the addition of certain metals like
molybdenum and niobium, which stabilizes the $\gamma$ phase at room
temperature and below \cite{Dholabhai07,Koelling73} In metallic U,
the three 5$f$ electrons of U hybridize with the 6$d$ and 7$s$
electrons, and show itinerant behaviors
\cite{Dholabhai07,Huda05,Gouder98}. Therefore, density functional
theories without any modifications concerning localized electronic
states are appropriate within a large extent to describe metallic U.

Recently, many theoretical and computational studies emerge on the
surface chemical properties of uranium and other actinide metal
surfaces. Using the density functional semicore pseudopotential
method, the chemisorption of carbon monoxide \cite{Dholabhai07} and
oxygen gases \cite{Huda05} on the $\gamma$-U surfaces have been
investigated focusing on the geometric, magnetic and electronic
properties of the system. The adsorption of carbon monoxide on
$\alpha$-U surfaces has also been studied using a plane-wave
ultrasoft pseudopotential \cite{Senanayake05}. For the interaction
between hydrogen and U, present theoretical studies are all carried
out on the $\alpha$-U surfaces \cite{Nie08,Taylor09,Taylor10}.
Different from the hydrogen dissociation on transition-metal
surfaces, where electronic hybridizations between hydrogen 1$s$
electrons and surface $d$ electrons induce the dissociation
\cite{Harris85}, the dissociation of hydrogen molecules on the
$\alpha$-U surface is because of 5$f$-1$s$ hybridizations
\cite{Nie08}, which is similar to the hydrogen dissociation on
$\delta$-Pu surfaces \cite{Huda05PRB}. Here in this paper, by using
first-principles calculations, we systematically study the
adsorption and dissociation processes of hydrogen molecules on the
$\gamma$-U(100) surface, and reveal that the $d$ electrons of U,
instead of its $f$ electrons, take part in the electronic
hybridizations with hydrogen 1$s$ electrons. This result clearly
presents different surface chemical properties of $\alpha$- and
$\gamma$-U. Considering that the $\gamma$ phase of U is available at
normal temperatures only after doping with 3$d$ transition metals,
here we also investigate the influences of surface Nb doping on the
dissociation properties of hydrogen molecules.

\section{Calculation Method}

Our calculations are performed within density functional theory
using the spin-polarized edition of Vienna {\it ab initio}
simulation package (VASP) \cite{VASP}. The Perdew-Burke-Ernzerhof
(PBE) \cite{PBE1,PBE2} generalized gradient approximation and the
projector-augmented wave potential \cite{PAW} are employed to
describe the exchange-correlation energy and the electron-ion
interaction, respectively. The cutoff energy for the plane-wave
expansion is set to 520 eV. The U(100) surface is modeled by a slab
composing of five atomic layers and a vacuum region of larger than
15 \AA~. The bottom two layers are fixed, and the other U layers are
free to move during the geometry optimizations of the uranium
surfaces. A 2$\times$2 supercell, in which each monolayer contains
four U atoms is adopted in the study of the H$_2$ adsorption. Our
test calculations have shown that the 2$\times$2 supercell is
sufficiently large to avoid the interaction between adjacent H$_2$
molecules. Integration over the Brillouin zone is done using the
Monkhorst-Pack scheme \cite{Monkhorst} with 5$\times$5$\times$1 grid
points. And a Fermi broadening \cite{Weinert1992} of 0.2 eV is
chosen to smear the occupation of the bands around the Fermi energy
$E_F$ by a finite-$T$ Fermi function and extrapolating to $T$=0 K.
The H$_2$ is placed on one side of the slab, namely, on the top
surface. The calculation of the potential-energy surface is
interpolated to 350 points with different bond length ($d_{\rm
H-H}$) and height of the mass center ($h_{\rm H_2}$) of H$_2$ at
each surface site. The calculated lattice constant of $\gamma$-U and
bond length of a free H$_2$ molecule are 3.43 and 0.75 \AA,
respectively, in good agreement with the experimental values of
3.467 \cite{Dholabhai07} and 0.74 \AA~ \cite{Huber1979}.

\section{Results and discussion}

\subsection{Hydrogen dissociation on the clean $\gamma$-U(100) surface}

The geometry and electronic properties of the clean $\gamma$-U(100)
surface is firstly investigated. After geometry optimization, we
find that the two topmost U layers relax significantly from the bulk
values. The first-second interlayer is contracted by 26.4\% and the
second-third interlayer is expanded by 15.6\%. The huge surface
relaxation reflects that the surface electronic structure should be
quite different from that in the bulk. The projected density of
states (PDOS) around the Fermi energies are then calculated for bulk
U and the $\gamma$-U(100) surface, and shown in Fig. 1(a). We can
see that in the PDOS for bulk U, the occupied electronic states are
itinerant without any localized peaks, which is in good agreement
with previous theoretical studies \cite{Dholabhai07,Huda05}. For the
U atom in the $\gamma$-U(100) surface, the narrow peaks in the
unoccupied $f$ states disappear. This result supports the
experimental observations in X-ray and ultraviolet photoelectron
spectroscopy, and auger electron spectroscopy that localization
effects are weak in U films \cite{Gouder98}. The PDOS for an U atom
in $\alpha$-U and in the $\alpha$-U(001) surface is also shown in
Fig. 1(b). One can see that there are more $f$ electrons for the U
atom in the $\alpha$-U(001) surface than in $\alpha$-U, which
corresponds to the charge transfer from bulk to the surface atomic
layer. Comparatively, such kind of charge redistribution is not seen
for the $\gamma$-U(100) surface.

After studying the geometry and electronic property of the
$\gamma$-U(100) surface, we build our models to calculate the
two-dimensional (2D) potential energy surface (PES) cuts for H$_{2}$
on the relaxed uranium surface. As shown in Fig. 2(a), there are
three different high-symmetry sites on the clean U surface,
respectively, the top, bridge (bri), and hollow (hol) sites. At the
bridge site, an adsorbed H$_{2}$ has three different high-symmetry
orientations, respectively along the $x$ (i.e., [$001$]), $y$ (i.e.,
[$010$]), and $z$ (i.e., [$100$]) directions. Nevertheless, at the
top and hollow sites, the $x$ and $y$ directions are degenerate, and
the three high-symmetry orientations become the $x$, $d$(i.e.,
[$011$]), and $z$ directions. Herein, we use top-$x,d,z$,
bri-$x,y,z$, and hol-$x,d,z$ to represent the nine high-symmetry
channels for the adsorption of H$_{2}$ on the $\gamma$-U(100)
surface, respectively. We have also constructed several low-symmetry
initial structures by rotating the H$_{2}$ molecule in the $XY$,
$YZ$, and $XZ$ planes with small angles. Through PES calculations,
we find that the energy barrier for H$_2$ dissociation along the
low-symmetry channels is always larger than along the high-symmetry
ones. Similar results have also been obtained for the
O$_{2}$/Pb(111) system where O$_2$ adsorption also prefers the
high-symmetry channels \cite{Yang08}. Therefore, we will only
discuss the obtained PES cuts along the high-symmetry channels.

From our PES calculations, we find that there are no molecular
adsorption states for H$_{2}$ on the $\gamma$-U(100) surface. The
calculated 2D PES cuts along the top-$x$, hol-$d$, bri-$y$ and
bri-$z$ channels are respectively listed in Figs. 3(a)-3(d). The PES
cuts along the top- and hol-$z$ channels have similar energy
distributions with the bri-$z$ channel, and thus are not listed.
Along the other bri-$x$, top-$x,d$ and hol-$x,d$ adsorption
channels, the H$_2$ molecule dissociates after overcoming small
energy barriers. Only in the PES cut along the hol-$d$ channel, we
find a local energy minimum after H$_2$ dissociation, which is shown
in Fig. 3(b). Along all the other dissociation channels, the H-H
bond length is larger than 2.4 \AA~ in the atomic adsorption states
after dissociation. So we do not see the local minima in the
calculated PES cuts. The local minimum point in Fig. 3(b)
corresponds to the adsorption state of two hydrogen atoms in the
same surface uranium square hollow. After geometry optimization from
this point, we find that the surface uranium atoms are distorted to
lower down the free electronic energy. And in the stable adsorption
state, the free energy of the adsorption system is 2.59 eV lower
than that of an isolated H$_2$ molecule plus a clean $\gamma$-U(100)
surface.

The most energetically favorable dissociation channel for H$_2$ is
along the bri-$y$ channel on the $\gamma$-U(100) surface, which is
found to be with no energy barriers. As shown in Fig. 3(c), the free
electronic energy of the adsorption system drops by about 0.4 eV at
the molecular height of $h_{\rm H_2}$=0.90 \AA, when the hydrogen
bond length $d_{\rm H-H}$ becomes 1.90 \AA. Our result that H$_2$
molecules dissociate without any energy barriers on the
$\gamma$-U(100) surface indicates different surface chemical
properties of $\gamma$-U with $\alpha$-U, because hydrogen
dissociation on the $\alpha$-U surface needs to overcome a small
energy barrier of 0.08 eV \cite{Nie08}. And as we will see in the
following, the electronic interactions during H$_2$ dissociation on
the $\gamma$-U surface is different from that on the $\alpha$-U
surface.

The PDOS evolution of H and U atoms along the bri-$y$ channel is
then analyzed to study the electronic interactions during the
barrierless dissociation process of H$_2$ on the $\gamma$-U(100)
surface. We have chosen four points along the minimum energy
dissociation path to calculate the PDOS, which are ($h_{\rm
H_2}$=1.17 \AA, $d_{\rm H-H}$=1.28 \AA), ($h_{\rm H_2}$=1.02 \AA,
$d_{\rm H-H}$=1.58 \AA), ($h_{\rm H_2}$=0.90 \AA, $d_{\rm H-H}$=1.90
\AA), and ($h_{\rm H_2}$=0.84 \AA, $d_{\rm H-H}$=2.38 \AA)
respectively. The obtained PDOS of the adsorption system at these
points are listed in Figs. 4(a)-(d) respectively. As shown in Fig.
4(a), at the molecular height of 1.17 \AA, a few electrons transfer
from the molecular orbital of H$_2$ to the unoccupied $d$ states of
U, which will be called as the charge donation process at following
discussions. At the same time, some electrons also transfer back
from U to H, forming a new peak in the hydrogen $s$ states near 2.25
eV below the Fermi energy, and we will call it as a charge
back-donation process. Since the change in free electronic energy is
still negligible at the molecular height of 1.17 \AA~, as shown in
Fig. 3(c), the above electronic hybridizations are not strong.

When the H$_2$ molecule further approaches the $\gamma$-U surface,
the electronic hybridizations become stronger, and the charge
donation and back-donation become more obvious. As shown in Figs.
4(b)-4(d), more and more electrons transfer from $s$ states of H to
the unoccupied $d$ states of U, and more and more electrons transfer
from $d$ and $f$ states of U back to hydrogen. We can also see from
Figs. 4(b)-4(d) that the electronic hybridization between H $s$
states and U $d$ states is always stronger than that with U $f$
states. And the charge donation process only happens between H and
$d$ electronic states of U. Therefore, $d$ electronic states of U
play very important roles during the dissociation of H$_2$ molecules
on the $\gamma$-U surface. These results are quite different from
the dissociation of H$_2$ molecules on the $\alpha$-U surface, where
$d$ electronic states of U is negligible, and the electronic
hybridization happens between hydrogen $s$ and uranium $f$
electronic states \cite{Nie08}.

To investigate the charge transfer for the dissociative adsorption
of H$_2$ on the $\gamma$-U(100) surface, we then calculate the
atomic charges for the final state of the minimum energy
dissociation path, using the Bader topological method \cite{Bader}.
It is found that after dissociation, the two hydrogen atoms together
gain 0.90 electrons, indicating that the ionic part of the H-U
bonding plays a significant role during the dissociation process
\cite{Nie08}. Different from the dissociation of H$_2$ on the
$\alpha$-U(001) surface, where the charge transfer happens only
between the topmost U atoms and hydrogen atoms \cite{Nie08}, here we
find that the topmost and second atomic layer of the $\gamma$-U(100)
surface both lose electrons to hydrogen atoms. In comparison with a
bare relaxed surface, the topmost and second layer loses 0.58 and
0.34 electrons respectively. This result indicates that the
electronic interaction between H$_2$ and the $\gamma$-U(100) surface
is not so localized as that between H$_2$ and the $\alpha$-U(100)
surface, and also prove the participation of more itinerant $d$
electronic states in interactions with hydrogen electrons.

\subsection{Hydrogen dissociation on the Nb-doped $\gamma$-U(100) surface}

Since the $\gamma$ phase of U is more stable after doping with 3$d$
transition metal atoms. We here also investigate the influences of
Nb-doping on the dissociation of H$_2$ molecules on the
$\gamma$-U(100) surface. Previous studies have already revealed that
doped Nb atoms thermodynamically prefer to substitute U atoms in
$\gamma$-U, rather than occupy octahedral or tetrahedral vacancies
\cite{Xiang08}. Thus we only consider the substitutional doping of
Nb atoms on the $\gamma$-U(100) surface.

Firstly, we do geometry optimization for the Nb-doped U surfaces,
with the Nb atom in the topmost, second, and third layer
respectively. To simplify our discussions, we will call them as the
UNb1, UNb2, and UNb3 surfaces in the following. The relaxed surface
structures of the UNb1 and UNb2 surfaces from the top view are shown
in Figs. 2(b) and 2(c) respectively. The adsorption sites of H$_2$
molecules on them are the same as that on the clean $\gamma$-U(100)
surface as depicted in Fig. 2(a). After geometry optimizations, the
uranium atoms in the doping layer are no longer in the same plane
with the doped Nb atoms. For example for the UNb1 surface, the $z$
coordinate of the Nb atom is 0.16 \AA~ larger than its nearest U
atoms, and 0.09 \AA~ smaller than its next neighboring U atoms. The
relative surface relaxation of Nb-doped $\gamma$-U surfaces can be
calculated by averaging the $z$ coordinates of the four atoms in the
same layer. In this way, the relative relaxation (i.e., $\Delta
d_{ij}/d_0$ with $d_{ij}$ and $d_0$ to be the interval between the
$i$th and $j$th atomic layer, and the lattice interval in bulk
$\gamma$-U) is calculated and summarized in Table I for the UNb1,
UNb2, and UNb3 surfaces. From the results listed in Table I, we can
see that the UNb1 surface has much smaller surface relaxations than
the clean $\gamma$-U(100) surface, indicating that their surface
electronic structures are different. As the Nb atom is doped deeper,
the surface relaxations tend to approach to the values of undoped
$\gamma$-U(100) surface.

The PDOS for the UNb1, UNb2, and UNb3 surfaces are also calculated
and shown in Figs. 5(b)-5(d), in comparison with the PDOS of the
undoped $\gamma$-U(100) surface shown in Fig. 5(a). We can see that
the electronic states around the Fermi energies are contributed by
both the $d$ electrons of Nb, and $d$, $f$ electrons of U.
Therefore, one can expect that the surface interactions with atoms
or molecules should be influenced by the doped Nb atoms, especially
for the UNb1 surface where the Nb atom is at the topmost layer. The
PES for H$_2$ on the UNb1 surface is then calculated to study the
influences.

The calculated 2D PES cuts for H$_2$ molecules on the UNb1 surface
along the top-$x$, hol-$d$, bri-$y$ and bri-$z$ channels are listed
in Figs. 6(a)-6(d) respectively. One can see great similarities in
the energy distributions with that on the clean $\gamma$-U(100)
surface. Firstly, along the bri-, top-, and hol-$z$ channels, H$_2$
molecules can hardly adsorb or dissociate. Secondly, the most
energetically favorable dissociation path is along the bri-$y$
channel, which is without any energy barriers. And at last, the
H$_2$ dissociation along the other channels needs to overcome small
energy barriers. Nevertheless, because of the introduction of a
surface Nb atom, there are also some new features. As shown in Figs.
3(a) and 6(a), the dissociation energy barrier is larger on top of
the Nb atom than on top of a U atom of the undoped U surface. And
for the most energetically favorable dissociation path, we see that
the free electronic energy lowers down much more quickly next to the
Nb atom than next to a U atom of the undoped U surface, as shown in
Figs. 3(c) and 6(c). These results indicate that the surface
electron distribution changes after doping with a Nb atom. The
larger energy reduction for H$_2$ dissociation on the UNb1 surface
also suggests that surface doping of Nb atoms reinforces H$_2$
dissociation, instead of hindering it.

The electronic interactions between H$_2$ and the UNb1 surface are
then studied for the most energetically favorable dissociation
channel bri-$y$. The electronic evolution is analyzed by calculating
the PDOS of H and U, Nb atoms at such four points: ($h_{\rm
H_2}$=1.50 \AA, $d_{\rm H-H}$=0.84 \AA), ($h_{\rm H_2}$=0.90 \AA,
$d_{\rm H-H}$=1.30 \AA), ($h_{\rm H_2}$=0.80 \AA, $d_{\rm H-H}$=1.53
\AA), and ($h_{\rm H_2}$=0.70 \AA, $d_{\rm H-H}$=2.12 \AA) along the
dissociation path, which are shown in Figs. 7(a)-7(d) respectively.
We can see from Fig. 7(a) that at the molecular height of 1.50 \AA,
no obvious electronic interactions happen between H and U, Nb atoms.
When the molecular height lowers to be 0.90 \AA, some electrons
transfer from H$_2$ to the unoccupied Nb $d$ electronic states,
corresponding to the charge donation process, and some other
electrons transfer back from $d$ states of U and Nb to the
antibonding orbital of H$_2$, corresponding to the charge
back-donation process. At the molecular height of 0.80 \AA, the
charge donation and back-donation processes become more obvious, as
shown in Figs. 7(e) and 7(f). We can also see from Fig. 7(e) that
the charge donation mainly goes to the $d$ orbitals around the Nb
atom. In comparison with the electronic interaction between H$_2$
and the undoped U surface, one can see that the $d$ electronic
states of Nb largely participate in both the charge donation and
back-donation processes. And the stronger electronic hybridizations
between hydrogen $s$ and Nb $d$ states lead to the fact that the
free electronic energy lowers down more quickly on the doped
$\gamma$-U surface.

\section{Conclusions}

In conclusion, we have systematically studied the dissociation of
H$_{2}$ molecules on the clean and Nb-doped $\gamma$-U(100)
surfaces. We find that both of the two reactions are with no energy
barriers. On the clean $\gamma$-U(100) surface, there are electronic
interactions between hydrogen electrons and U $d$ electrons, which
cause that H$_2$ molecules dissociate without any energy barriers
along the bri-$y$ channel. This mechanism is quite different from
the H$_2$ dissociation on the $\alpha$-U surfaces, where $f$
electrons of U, instead of $d$ electrons take part in the electronic
interactions. After surface doping of a Nb atom, we find that not
only the H$_2$ dissociation is with no energy barriers, but also the
dissociated hydrogen atoms bond stronger with the surface. The $d$
electronic states of Nb participate in the electronic interactions
with hydrogen, and causes the larger free energy reduction.

\begin{acknowledgments}
This work was supported by NSFC under Grants No. 10904004 and No.
90921003, and Foundations for Development of Science and Technology
of China Academy of Engineering Physics under Grants No.
2011B0301060, No. 2011A0301016, and No. 2008A0301013.
\end{acknowledgments}

\clearpage
\begin{table}[ptb]
\caption{The relative surface relaxation for the clean
$\gamma$-U(100) surface, and the UNb1, UNb2, UNb3 surfaces. $d_{ij}$
represents for the interval between the $i$th and $j$th atomic layer
of each surface, and $d_0$ is the lattice interval along the [100]
direction of bulk $\gamma$-U.}
%\centering%
\begin{tabular}
[c]{ccccc}\hline\hline
~~~surfaces~~~ & ~~~$\gamma$-U(100)~~~ & ~~~UNb1~~~ & ~~~UNb2~~~ & ~~~UNb3~~~ \\
\hline
$\Delta d_{23}/d_0$ & 26.4\% & 17.4\% & 23.8\% & 29.9\% \\
$\Delta d_{34}/d_0$ & 15.6\% & 7.0\%  & 21.9\% & 16.0\% \\
$\Delta d_{45}/d_0$ & 9.8\%  & 0.7\%  & 10.4\% & 11.1\% \\
\hline
\end{tabular}\label{deltd}
\end{table}
\clearpage

\noindent\textbf{List of captions} \\

\noindent\textbf{Fig.1}~~~ (Color online) The projected density of
states for a uranium atom in bulk and in the (100) surface of
$\gamma$- (a) and $\alpha$-U (b). The Fermi energies are set to be
zero. \\

\noindent\textbf{Fig.2}~~~ (Color online) (a) Top view of the clean
$\gamma$-U(100) surface with the high-symmetry adsorption sites and
high-symmetry H$_2$ orientations depicted. (b) and (c) Top view of
the UNb1 [U(100) surface with a doping Nb atom in the topmost layer]
and UNb2 [U(100) surface with a doping Nb atom in the second layer]
surfaces. (d) Side view of the adsorption model for H$_2$ on the
clean $\gamma$-U(100) surface. Blue, grey, and pink balls represent
for hydrogen, uranium, and niobium atoms respectively. The adopted
supercell is depicted by dashed lines. \\

\noindent\textbf{Fig.3}~~~ (Color online) The 2D PES cuts for the
adsorption of hydrogen molecules along the (a) top-$x$, (b) hol-$d$,
(c) bri-$y$, and (d) bri-$z$ channels on the clean $\gamma$-U(100)
surface. The total energy of a free H$_2$ molecule plus that of the
$\gamma$-U(100) surface is set to be the energy zero. \\

\noindent\textbf{Fig.4}~~~ (Color online) The projected density of
states for the H$_2$/$\gamma$-U(100) adsorption system along the
energetically most favorable dissociation path with the height of
H$_2$ center of mass to be 1.17 \AA~ (a), 1.02 \AA~ (b), 0.90 \AA~
(c), and 0.84 \AA~ (d). The Fermi energies are all set to be zero.
(e) and (f) The partial charge density distributions for the two
energy peaks denoted in (c). Blue and grey balls represent for
hydrogen and uranium atoms respectively. The yellow dots represents
for the isosurface of partial charge density. \\

\noindent\textbf{Fig.5}~~~ (Color online) (a) The projected density
of states for the clean $\gamma$-U(100) surface. (b), (c), and (d)
The projected density of states for the Nb doped $\gamma$-U(100)
surface, with the Nb atom at the topmost, second, and third layer.
The Fermi energies are all set to be zero. \\

\noindent\textbf{Fig.6}~~~ (Color online) The 2D PES cuts for the
adsorption of hydrogen molecules along the (a) top-$x$, (b) hol-$d$,
(c) bri-$y$, and (d) bri-$z$ channels on the UNb1 surface. The total
energy of a free H$_2$ molecule plus that of the Nb doped
$\gamma$-U(100) surface is set to be the energy zero. \\

\noindent\textbf{Fig.7}~~~ (Color online) The projected density of
states for the adsorption system of H$_2$ on the Nb-doped
$\gamma$-U(100) surface, along the energetically most favorable
dissociation path with the height of H$_2$ center of mass to be 1.50
\AA~ (a), 0.90 \AA~ (b), 0.80 \AA~ (c), and 0.70 \AA~ (d). The Fermi
energies are all set to be zero. (e) and (f) The partial charge
density distributions for the two energy peaks denoted in (c). Blue,
grey, and pink balls represent for hydrogen, uranium, and niobium
atoms respectively. The yellow dots represents for the isosurface of
partial charge density. \\

\clearpage

\begin{figure}
\includegraphics[width=1.0\textwidth]{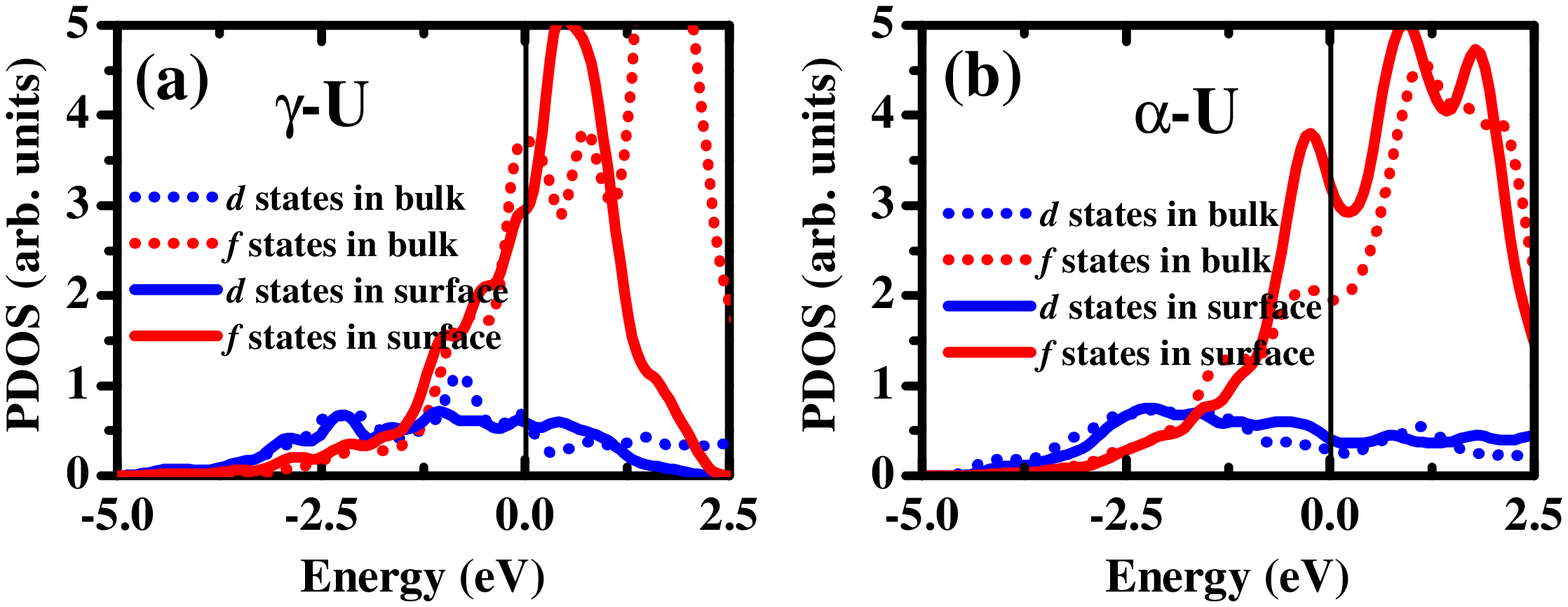}
\caption{\label{fig:fig1}}
\end{figure}
\clearpage
\begin{figure}
\includegraphics[width=1.0\textwidth]{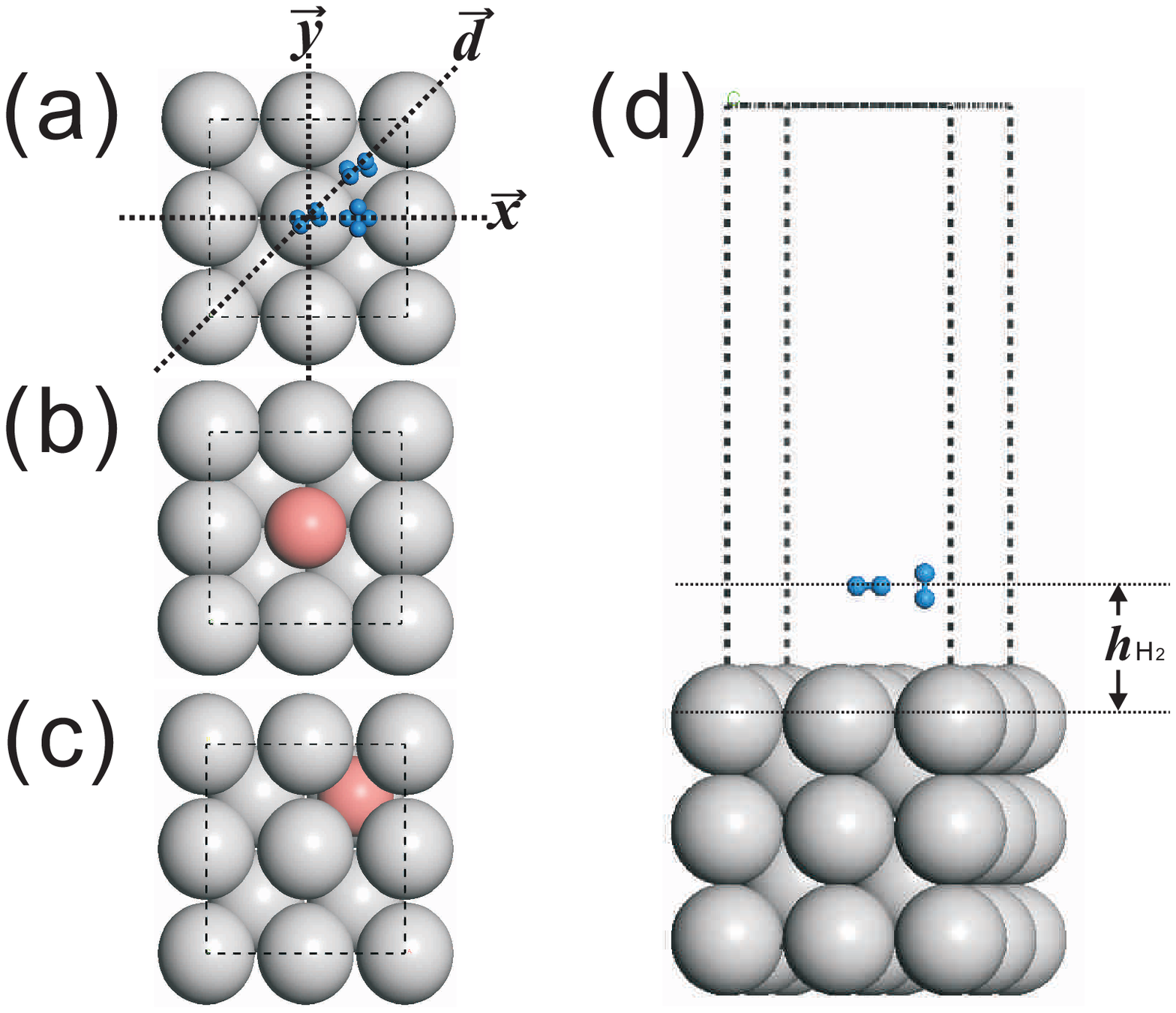}
\caption{\label{fig:fig2}}
\end{figure}
\clearpage
\begin{figure}
\includegraphics[width=1.0\textwidth]{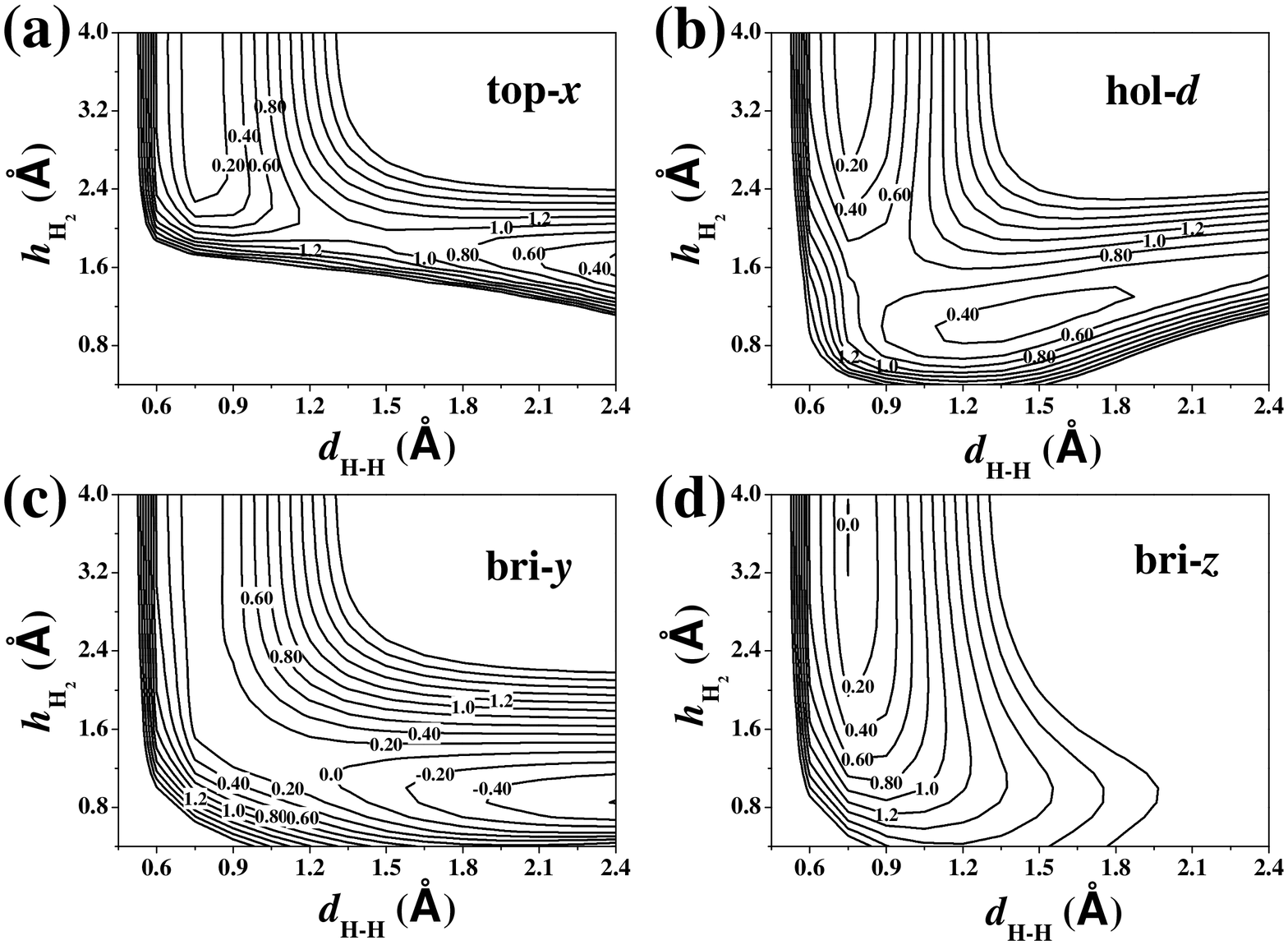}
\caption{\label{fig:fig3}}
\end{figure}
\clearpage
\begin{figure}
\includegraphics[width=1.0\textwidth]{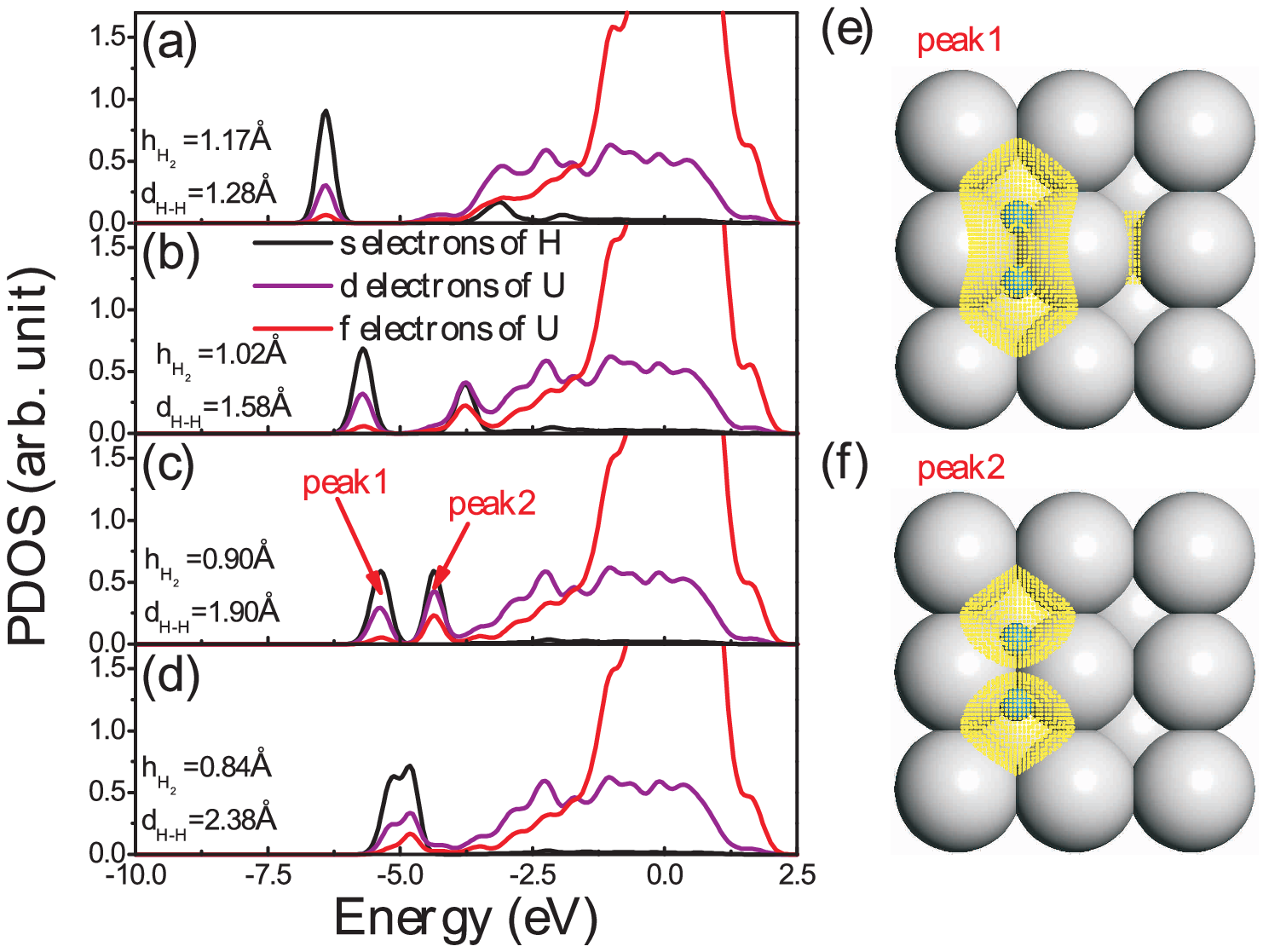}
\caption{\label{fig:fig4}}
\end{figure}
\clearpage
\begin{figure}
\includegraphics[width=1.0\textwidth]{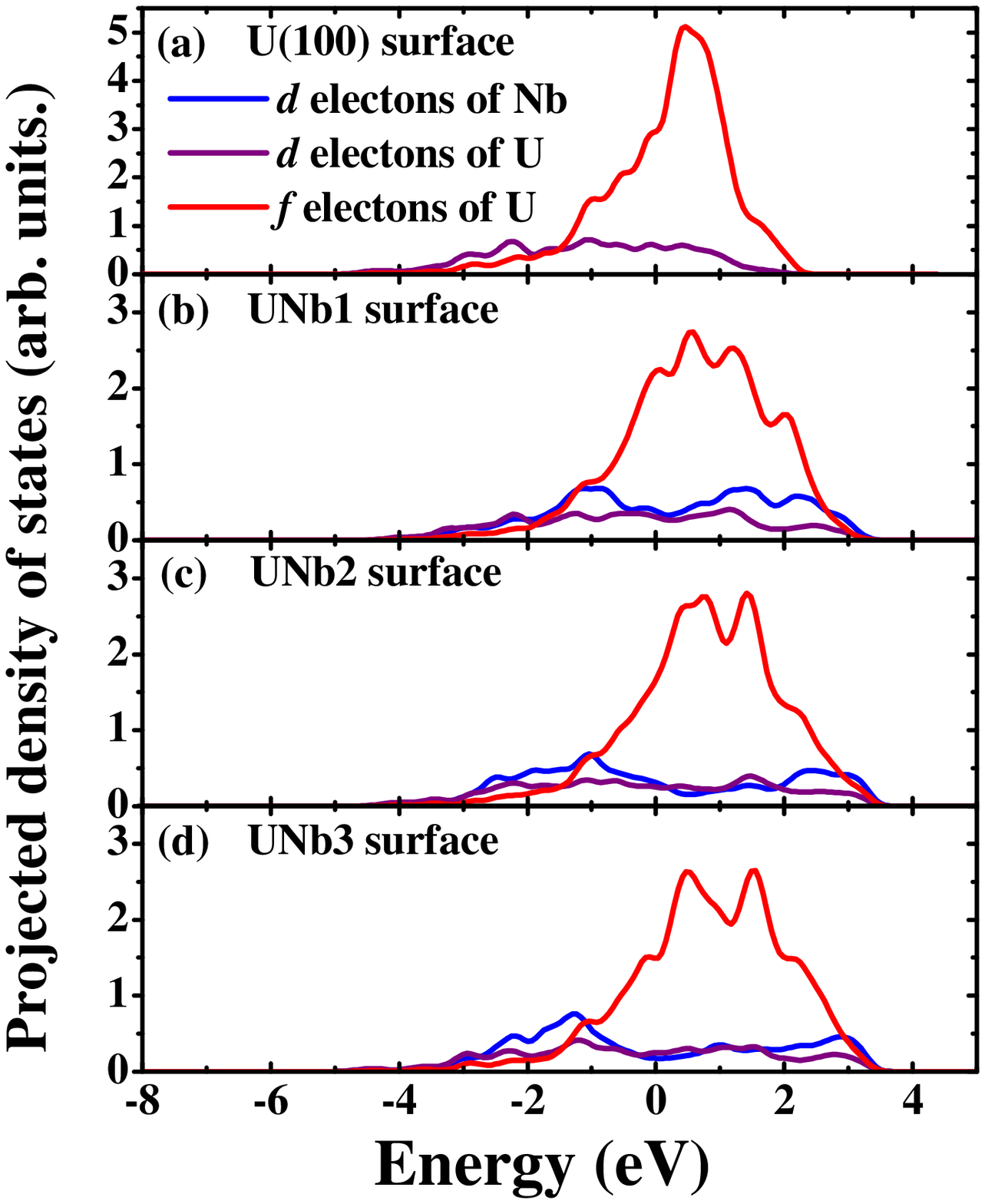}
\caption{\label{fig:fig5}}
\end{figure}
\clearpage
\begin{figure}
\includegraphics[width=1.0\textwidth]{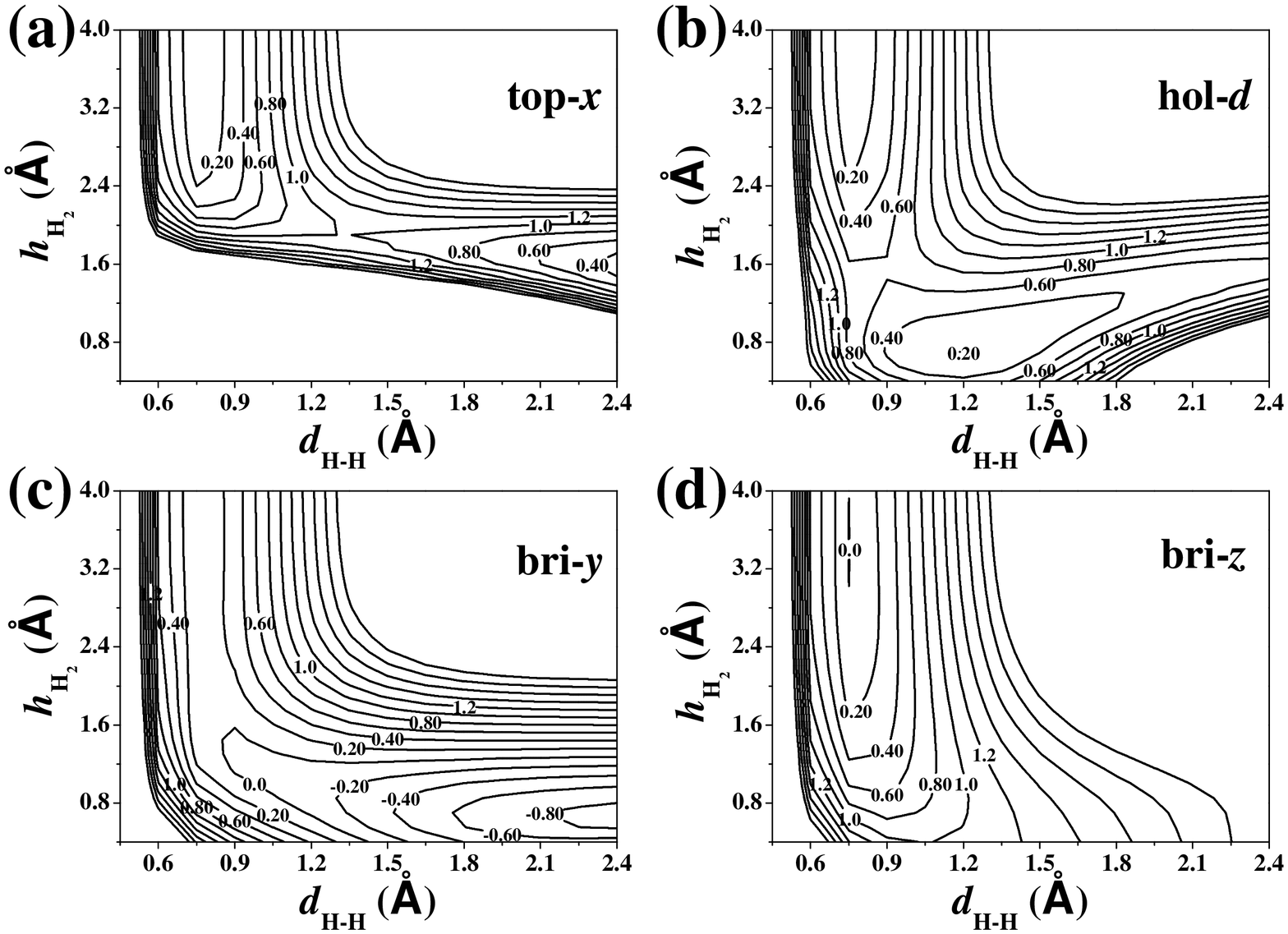}
\caption{\label{fig:fig6}}
\end{figure}
\clearpage
\begin{figure}
\includegraphics[width=1.0\textwidth]{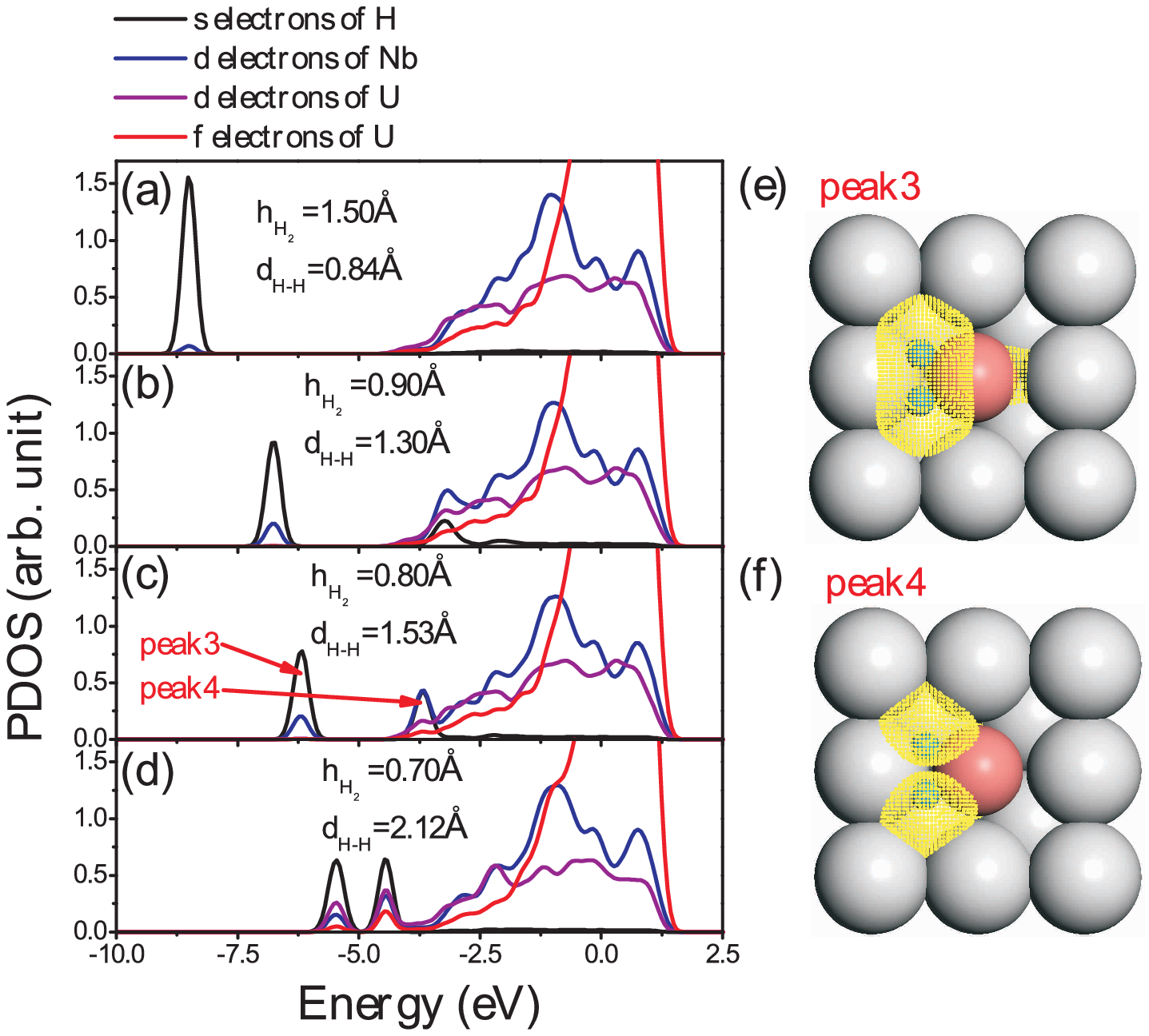}
\caption{\label{fig:fig7}}
\end{figure}
\end{document}